\documentclass[a4paper,12pt]{article}
\pdfoutput=1
\usepackage[margin=1in]{geometry}
\usepackage{graphicx,xcolor}
\usepackage{mathtools,braket,blkarray}
\allowdisplaybreaks
\usepackage{amssymb,amsfonts,bbm,relsize}
\usepackage[width=0.9\textwidth,footnotesize]{caption}
\usepackage[font=scriptsize]{subcaption}
\usepackage{cite,hyperref,url}
\hypersetup{pageanchor=false}
\usepackage{booktabs,multirow,makecell}
\usepackage[utf8]{inputenc}
\usepackage{cleveref}
\usepackage{amsmath}
\usepackage{ulem}

\newcommand{\be}{\begin{equation}}
\newcommand{\ee}{\end{equation}}
\newcommand{\ba}{\begin{eqnarray}}
\newcommand{\ea}{\end{eqnarray}}

\begin{document}

\begin{titlepage}
\begin{center}
{\bf\Large Leptogenesis in Realistic Flipped SU(5) 
  } \\[12mm]
Stephen~F.~King$^{a}$~\footnote{E-mail: \texttt{king@soton.ac.uk}},
George~K.~Leontaris$^{b}$~\footnote{E-mail: \texttt{leonta@uoi.gr}},
Luca~Marsili$^{c}$~\footnote{E-mail: \texttt{luca.marsili@ific.uv.es}}, and
Ye-Ling~Zhou$^{d}$~\footnote{E-mail: \texttt{zhouyeling@ucas.ac.cn}}
\\[-2mm]

\end{center}
\vspace*{0.50cm}
\centerline{$^{a}$ \it
School of Physics and Astronomy, University of Southampton,}
\centerline{\it
SO17 1BJ Southampton, United Kingdom }
 \vspace*{0.2cm}
\centerline{$^{b}$~\it
			Physics Department, University of Ioannina, 45110, Ioannina, 	Greece}
 \vspace*{0.2cm}
\centerline{$^{c}$~\it
			Instituto de F\`isica Corpuscular (IFIC), Universitat de Val\`encia, }
   \centerline{\it Parc Cientific UV, C/ Catedratico Jose Beltran 2, E-46980 Paterna, Spain}
 \vspace*{0.2cm}
\centerline{$^{d}$ \it School of Fundamental Physics and Mathematical Sciences,}
\centerline{\it Hangzhou Institute for Advanced Study, UCAS, Hangzhou 310024, China}
\vspace*{1.20cm}

\begin{abstract}
{\noindent

We study thermal leptogenesis in realistic supersymmetric flipped $SU(5)\times U(1)$  unification. As up-type quarks and neutrinos are arranged in the same multiplets, they exhibit strong correlations, and it is commonly believed that the masses of right-handed (RH) neutrinos are too hierarchical to fit the low-energy neutrino data. This pattern generally predicts a lightest RH neutrino too light to yield successful leptogenesis, with any lepton-antilepton asymmetry generated from heavier neutrinos being washed out unless special flavour structures are assumed. We propose a different scenario in which the lightest two RH neutrinos $N_1$ and $N_2$ have nearby masses of order $10^9$ GeV, with thermal leptogenesis arising non-resonantly from both $N_1$ and $N_2$. We show that this pattern is consistent with all data on fermion masses and mixing and predicts the lightest physical left-handed neutrino mass to be smaller than about $10^{-7}$~eV. The Dirac phase, which does not take the maximal CP-violating value, plays an important role in leptogenesis. 
}
\end{abstract}
\end{titlepage}

\section{Introduction}

The flipped $SU(5)\times U(1)$ Grand Unified Theory (GUT) model~\cite{Barr:1981qv,Derendinger:1983aj} is a compelling alternative to the Georgi-Glashow (standard) $SU(5)$ GUT~\cite{Georgi:1974sy}. It exhibits a number of interesting features that are essential in addressing several unsettling issues present in (at least in  minimal versions of) the standard SU(5) framework. In this respect, the  most notable characteristic of the flipped SU(5) model is that the standard hypercharges and the electric charges for the Standard Model fermions emanate from a different $U(1)_Y$ hypercharge embedding. This new hypercharge  arrangement is obtained by associating $U(1)_Y$  with a linear combination of the $U(1)_y\subset SU(5)$ and a second $U(1)_{\chi}$ which essentially corresponds to the Abelian factor embedded in  $SO(10)$.
This modification of the original $SU(5)$ leads to far reaching theoretical and phenomenological implications. Under the new assignment, quarks and leptons are distributed  differently in the SU(5) representations, and moreover the right-handed (RH) neutrinos are also integrated in the spectrum, which generate  naturally light neutrino masses through an effective seesaw mechanism. Furthermore, 
the symmetry breaking to the Standard Model is achieved only with the fundamental ${\bf 10}+\overline{\bf 10}$ Higgs representations, as opposed to the standard Georgi-Glashow model where the adjoint (i.e. the ${\bf 24}$-plet) is required for the $SU(5)$  symmetry breaking. By virtue of this property, the flipped $SU(5)$ version can be elegantly embedded within a superstring theory framework (for example, such as in the d-4 fermionic formulation~\cite{Antoniadis:1987dx} of the heterotic string theory, and a version~\cite{King:2010mq} in F-theory), providing an ultraviolet (UV) completion  and  thus, a connection with quantum gravity at high energy scales. Moreover, 
the above novel features have made the model attractive for
 addressing successfully challenging phenomenological issues. Thus, for instance, it can naturally provide an interpretation to the neutrino oscillations and also, avoid fast proton decay.  For all those merits, since its early days the flipped $SU(5)$ model is an evolving area of research, remaining a compelling candidate for physics beyond the Standard Model until today. Due to its UV completion, it is natural to assume that the flipped $SU(5)$ respects supersymmetry (SUSY), and we shall do so here.

	In a previous paper~\cite{King:2023wkm}, aiming to further refine and enhance the predictability of flipped SU(5) SUSY GUTs,  we have made a detailed investigation and considered its broader implications for particle physics and cosmology.  Thus, we first performed a renormalisation group analysis to settle the unification scale $M_{\rm GUT}$ and other related high mass scales of the $SU(5)\times U(1)_{\chi}$ theory.  This determined 
	the GUT scale $M_{\rm GUT} \gtrsim 10^{16}$ GeV, while it was found that the 	$U (1)_{\chi}$  breaking scale (associated with $B-L$),  can stretch  over a wide  range of scales without 
	having significant impact on $M_{\rm GUT}$.  Subsequently, we investigated two viable scenarios of  fermion masses and mixings and derived a light neutrino spectrum compatible with the present neutrino data. Next,
	we computed the contributions to proton decay from dimension-five  and dimension-six operators, while we found that the former can be adequately  suppressed by virtue of the missing partner mechanism in this model. In general, contribution to the partial decay width $p \to \pi^+\bar\nu$  are highly suppressed, while the dominant channel in this model  is $p \to  \pi^0e^+$. 
Further, a mechanism of generating cosmic strings associated with the $U(1)_{\chi}$ (hence, with the $B-L$ scale) is effective in this model.  These (metastable) cosmic strings can provide an interpretation to the recently observed NANOGrav  stochastic
gravitational wave background \cite{NANOGrav:2023gor,NANOGrav:2023hvm}. 
	
	While many phenomenological aspects of flipped $SU(5)$ have been examined in detail  over the last decades, the leptogenesis scenario has not received much attention~\footnote{Note that a recent paper on flipped $SU(5)$ leptogenesis generates RH neutrino masses via a two loop mechanism~\cite{Fonseca:2023per}. However such a mechanism is not consistent with SUSY as assumed here.}, hence, in the present work  we  address this issue  in some detail.  Leptogenesis, is an attractive scenario for  generating  the observed  baryon asymmetry of the Universe, however, it is also  a dynamical mechanism on its own right, since it can make prediction for the leptonic sector as well.  The
	leptogenesis can in principle be realised if heavy RH neutrinos are present, therefore, flipped $SU(5)$ is a suitable candidate since it is the minimal GUT incorporating the RH neutrinos in its spectrum. 	In the present work we rely on our previous construction~\cite{King:2023wkm} of the flipped model to investigate the leptogenesis scenario.	We start by performing a detailed analysis of the fermion masses and mixing and determine regions of the parameter space where leptogenesis  is successfully implemented.  This requires a RH neutrino  mass spectrum  that differs from the strong hierarchical case of  standard scenarios presented in most of the previous investigations in the framework of the flipped $SU(5)$. Hence, in the present work we constrain the parametric space in the  region where  the two lightest RH neutrinos are nearly degenerate while the third RH eigenstate is much heavier, i.e., $M_1 \approx M_2 \ll M_3$. This case can avoid the strong restriction of $N_2$ leptogenesis, where $N_1$ is too light to generate enough lepton asymmetry and the lepton asymmetry generated by $N_2$ should be carefully reserved to avoid the washout by $N_1$ \cite{DiBari:2010ux,DiBari:2015oca,DiBari:2015svd,DiBari:2017uka}. In our scenario, both $N_1$ and $N_2$ have masses higher than $10^9$~GeV.  Using publicly available packages, we calculate the baryon asymmetry using the density matrix formalism and determine the specific conditions and the parameter region where leptogenesis can be achieved.

The layout of the remainder of this paper is as follows.	In section \ref{sec:flavour} we present the spectrum of the model and describe the  breaking  pattern of the (flipped) $SU(5)\times U(1)_{\chi}$ symmetry. In section \ref{sec:flavour},  we perform an  analytical and numerical investigation of the fermion mass textures and mixing, under the assumption that the two lightest RH neutrinos are nearly degenerate. More details on the analytical derivation of this flavour texture is given in the appendix. In section \ref{sec:leptogenesis}, we analyse the implications of the derived fermion mass spectrum of the previous section and constrain the available parametric space to  achieve a successful  leptogenesis scenario. In section \ref{sec:conclusion} we present our conclusions.

\section{Flipped SU(5) and  the Fermion masses \label{sec:flavour}}

	\begin{table}[h!]
		\begin{center}
			\begin{tabular}{|c | c | c | l |}
				\hline \hline
				& Superfields & SM decomposition & Role in the model
				\\\hline
				&$F=({\bf 10},-\frac{1}{2})$ & ( $Q, d^c, \nu^c$) & \\
				Matters &$\bar f=(\bar{\bf 5}, +\frac{3}{2})$ &   ($u^c, L$) & SM matters \& RH neutrinos \\
				&$e^c=({\bf 1}, -\frac{5}{2})$ &  $e^c$ & \\\hline
				&$h=({\bf 5}, +1)$ & $(D, h_d)$ & {Generate Dirac  fermion masses} \\
				&$\bar h=(\overline{\bf 5}, -1)$ &  $(\bar D, h_u)$ & {for leptons and up,down quarks}\\\cline{2-4}
				Higgses &$H=({\bf 10}, -\frac12)$ &  ( $Q_H, d_H^c, \nu_H^c$) & {Generate $\nu^c$ mass and} \\
				& $\bar H=(\overline{\bf 10},{+\frac 12})$ &( $\bar Q_H, \bar d_H^c, \bar\nu_H^c$) & {trigger $U(1)_{B-L}$ breaking} \\\cline{2-4}
				& $\Sigma=({\bf 24},0)$ & - & Triggers the breaking of $SU(5)$ \\		\hline \hline
			\end{tabular}
		\end{center}
		\caption{$SU(5)\times U(1)_{\chi} $  representations for matter and Higgs fields of our $SU(5)\times U(1)_{\chi} $ GUT model and their role in symmetry breaking. Standard Model hypercharge is identified as
  $Y=-\frac{1}{5}(y+2\chi)$, where $y$ is the generator associated with the $U(1)_y\subset SU(5)$. }\label{tab:flipped_content}
	\end{table}

We briefly review the fermion masses and mixing  in the flipped  $SU(5)$ model, which was outlined in our former paper~\cite{King:2023wkm}. There are three matter multiplets, $({\bf10},{-\frac 12 })_i$, $(\bar{{\bf5}},{\frac 32})_i$ and $({\bf1},{-\frac 52})_i$ in the gauge symmetry $SU(5)\times U(1)_{\chi}$, where the index $i=1,2,3$ takes the values  in the flavour space. A copy of quintet Higgses, $({\bf5},{1})$ and $(\bar{{\bf5}},{-1})$, are included  for the $SU(5)\times U(1)_{\chi}$ invariant Yukawa couplings with fermions. Additional Higgses are necessary to trigger the spontaneous breaking of the GUT symmetry and intermediate symmetries above the electroweak scale. As they are not crucial for fermion masses, they will not be reviewed in this work, and we refer to \cite{King:2023wkm} for those details. With matter and Higgs field introduced,  Yukawa couplings consistent with the GUT symmetry can be constructed. The charged fermions masses in particular arise via the following superpotential terms
{\begin{eqnarray}
\label{eq:FermCoupl}
{\cal W}_d&=&(Y_d)^*_{ij}\,F_i  \, F_j \, h \;\to \; (Y_d)^*_{ij}\, Q_i\,d^c_j\,h_d \,, \nonumber\\
{\cal W}_u&=& (Y_u)^*_{ij}\, F_i \, \bar{f}_j \, \bar{h} \,\;\to \; (Y_u)^*_{ij}\, [ Q_i\,u^c_j +\nu^c_i\,L_j ]\, h_u \,, \nonumber\\
{\cal W}_l&=&(Y_l)^*_{ij}\;e^c_j \, \bar{f}_i \, h \,\;\;\to \; (Y_l)^*_{ij}\, e^c_j\,L_i \,h_d \,.
\end{eqnarray}
}Here $Y_u$, $Y_d$ and $Y_l$ are $3\times 3$ Yukawa matrices and $Y_d$ is symmetric. These coefficient matrices are introduced with a complex conjugation to match with the SM left-right  non-SUSY convention. 
The field arrangement requires the Yukawa coupling matrices  satisfying
\begin{eqnarray}
Y_d^T = Y_d \,,\quad 
Y_u = Y_\nu^T\,.
\end{eqnarray} 
In particular, the Dirac Yukawa coupling matrix $Y_\nu$ is correlated with the up-quark Yukawa coupling, inheriting the hierarchical structure of the latter.
Majorana masses  for the RH neutrinos can be obtained via a higher order term
\begin{eqnarray}
{\cal W}_{\nu^c}&=&(\lambda^{\nu^c})^*_{ij}\frac{1}{2M_S}\,{\bar H}\,{\bar H}\,{F_i}\,F_j\to \frac12 (M_{R})_{ij}^*\nu^c_i\nu^c_j \,.
\label{Maj}
\end{eqnarray}
where $(M_R)_{ij} =( \lambda^{\nu^c})_{ij} \frac{\langle \bar\nu^c_H\rangle^2}{M_S}$. The light neutrinos gain masses via the usual type-I seesaw mechanism,
\begin{eqnarray}
M_{\nu} = - Y_\nu M_{R}^{-1} Y_\nu^T v_u^2 \,,
\end{eqnarray}
where we takes the Higgs VEV $v_u = \langle h_u \rangle \simeq 175~{\rm GeV}$ for $\tan \beta \gg 1$. 

{We further check if there would be additional dim-4 superpotential terms (i.e., dim-5 operators) contributing to fermion masses. Those invariant under $SU(5)$ and $U(1)_\chi$ are
\begin{eqnarray}
(F_i \bar{f}_j)_{\bf 5} (H H)_{\bar{\bf 5}_S}, \quad 
(F_i \bar{f}_j)_{\bf 5,45} (\bar{h} \Sigma)_{\bf 5, 45}, \quad
(F_i F_j)_{\bf 5,45} (h \Sigma)_{\bf 5, 45}.
\end{eqnarray}
The first operator vanishes at the VEV of $H$ and thus has no contribution to fermion masses. The second and third ones, after $\Sigma$ gains the VEV at the GUT scale and $h$ gains VEV at the EW scale, give contributions to up-type quark, neutrino masses and down-type quark, respectively. These terms can be forbidden by introducing a parity symmetry \cite{Antoniadis:1987dx}, here more precisely, $\Sigma \to -\Sigma$.}

All Yukawa mass matrices, can be diagonalised as 
\begin{eqnarray}
&&U_f^\dag Y_f U_f' = \hat{Y}_f \equiv {\rm diag}\{ y_{f1}, y_{f2}, y_{f3} \} \,, \nonumber\\
&&U_\nu^\dag M_\nu U_\nu^* = \hat{M}_\nu \equiv {\rm diag} \{ m_1, m_2, m_3\} \,, \nonumber\\
&&U_{R}^\dag M_{R} U_{R}^* = \hat{M}_R \equiv {\rm diag} \{ M_1, M_2, M_3\} \,,
\end{eqnarray}
where $f= u, d, e$, and $( y_{u1}, y_{u2}, y_{u3} ) = ( y_{u}, y_{c}, y_{t} )$, and etc. After the diagonalisation, we obtain the quark and lepton flavour mixing matrices as
$V_{\rm CKM} = U_u^\dag U_d$ and 
$U_{\rm PMNS} = U_l^\dag U_\nu$. 
In particular, the lepton flavour mixing matrix, i.e., the PMNS matrix, up to three unphysical phases on the left hand side, is parametrised as follows
\begin{eqnarray}
U_{\rm PMNS} = P_l
	\left(\begin{matrix}
		c^{}_{12} c^{}_{13} & s^{}_{12} c^{}_{13} &
		s^{}_{13} e^{-{ i} \delta} \cr \vspace{-0.4cm} \cr
		-s^{}_{12} c^{}_{23} - c^{}_{12}
		s^{}_{13} s^{}_{23} e^{{ i} \delta} & c^{}_{12} c^{}_{23} -
		s^{}_{12} s^{}_{13} s^{}_{23} e^{{ i} \delta} & c^{}_{13}
		s^{}_{23} \cr \vspace{-0.4cm} \cr
		s^{}_{12} s^{}_{23} - c^{}_{12} s^{}_{13} c^{}_{23}
		e^{{ i} \delta} &- c^{}_{12} s^{}_{23} - s^{}_{12} s^{}_{13}
		c^{}_{23} e^{{ i} \delta} &  c^{}_{13} c^{}_{23} \cr
	\end{matrix} \right)\, P_\nu \,,
 \label{eq:xxx}
\end{eqnarray}
where $\theta_{ij}$ (for $ij = 12,13,23$) are three mixing angles, $\delta$ is the Dirac CP phase, $P_\nu = {\rm diag}\{1,e^{i\alpha_{21}/2},e^{i \alpha_{31}/2}\}$ is the Majorana phase matrix and $P_l = {\rm diag} \{ e^{i \beta_1}, e^{i \beta_2}, e^{i \beta_3}\}$ is a diagonal phase matrix without physical correspondence at low energy. The CKM matrix can be parameterised similarly to Eq.~\eqref{eq:xxx} with a $3\times 3$ matrix in the middle, involving three mixing angles $\theta_{ij}^q$ and a Dirac CP phase $\delta^q$, accompanied with two diagonal phase matrices $P_u$ and $P_d$ on both sides. However, as quarks are all Dirac fermions, the two phase matrices are unphysical at low energy. 

Without loss of generality, we make a basis rotation to the basis where $U_u = U_u'= U_l' = 1$. This is done by performing $3\times 3$ unitary transformations for $({\bf 10}, -\frac12)$, $(\bar{\bf5},\frac32)$, $({\bf1},-\frac52)$ in their flavour space, respectively. Since  $Y_d$ is symmetric, $U_d' = U_d^*$ is satisfied. Then, we arrive at
\begin{eqnarray} \label{eq:master}
Y_u &=& Y_\nu = \hat{Y}_u\,,\nonumber\\
Y_d &=& V_{\rm CKM} \hat{Y}_d Y_{\rm CKM}^T \,, \nonumber\\
Y_l &=& U_\nu \, U_{\rm PMNS}^\dag \, \hat{Y}_l \,, \nonumber\\
M_\nu &=& U_\nu \hat{M}_\nu U_\nu^T \,, \nonumber\\
M_{R} &=& \hat{Y}_u \, U_\nu^* \hat{M}_\nu^{-1} U_\nu^\dag \, \hat{Y}_u v_u^2~.
\end{eqnarray}
In this basis, $\hat{Y}_f$ and $\hat{M}_\nu$ are fixed by the corresponding quark masses, $U_{\rm CKM}$ and $U_{\rm PMNS}$ are determined by experimental data of quark mixing and lepton mixing (up to the two unknown Majorana phases). The main undetermined part is  the unitary matrix $U_\nu$. 

We have discussed two extreme cases with regard to $U_\nu$ in the former paper~\cite{King:2023wkm}, namely:
\begin{itemize}
\item[S1)] $U_\nu = U_{\rm PMNS}$, i.e., $U_l = U_\nu U_{\rm PMNS}^\dag = {\bf 1}$. The Yukawa  mass matrices are simplified to
\begin{eqnarray}
Y_l &=& \hat{Y}_l \,, \nonumber\\
M_\nu &=& U_{\rm PMNS} \hat{M}_\nu U_{\rm PMNS}^T \,, \nonumber\\
M_R &=& \hat{Y}_u U_{\rm PMNS}^* \hat{M}_\nu^{-1} U_{\rm PMNS}^\dag \hat{Y}_u v_u^2~.
\end{eqnarray}

\item[S2)] $U_\nu = \mathbf{1}$, i.e., $U_l = U_\nu U_{\rm PMNS}^\dag = U_{\rm PMNS}^\dag$.
\begin{eqnarray}
Y_l &=& U_{\rm PMNS}^\dag \, \hat{Y}_l \,, \nonumber\\
M_\nu &=& \hat{M}_\nu \,, \nonumber\\
M_R &=& \hat{Y}_u \hat{M}_\nu^{-1} \hat{Y}_u v_u^2 \,.
\end{eqnarray}
\end{itemize} 
Both cases give too hierarchical mass spectrum of RH Neutrinos, $M_1: M_2: M_3 \propto m_u^2: m_c^2: m_t^2$. In particular, the lightest one acquires a mass $M_1 \sim m_u^2 / m_\nu < 10^{6}$~GeV, which cannot provide a source to generate enough lepton-antilepton asymmetry to address the matter-antimatter problem. The second case S2), which is even worse as we have confirmed, gives no CP violation for the RH neutrino decay. 

In the following sections, we will discuss how to use leptogenesis as a criterion to pick up leptogenesis-favoured $U_\nu$ and the corresponding flavour patterns.

\section{The flavour pattern} \label{sec:flavour}

We make the following assumptions for the RH neutrino mass matrix $M_R$: 
\begin{itemize}
\item The two light RH neutrinos are assumed to have nearly degenerate masses, and much lighter than the heaviest one, i.e.,
\begin{eqnarray}
M_1 \simeq M_2 \ll M_3 \,.
\end{eqnarray}
This allows us to perform the following parametrisation 
\begin{eqnarray}
M_{1,2} = M(1\mp\delta_M) \,,\quad
M_3 = M / \kappa~.
\end{eqnarray}
Then we can approximate the inverse of $M_R$ as
\begin{eqnarray} \label{eq:inv_M_R}
M_R^{-1} &\simeq& 
\frac{1}{M} \left\{
U_R \left(
\begin{array}{ccc}
 1 & 0 & 0 \\
 0 & 1 & 0 \\
 0 & 0 & 0 \\
\end{array}
\right) U_R^T
+
U_R \left(
\begin{array}{ccc}
 \delta_M & 0 & 0 \\
 0 & -\delta_M & 0 \\
 0 & 0 & \kappa \\
\end{array}
\right) U_R^T
\right\} \,,
\end{eqnarray}
where $U_R$ is the unitary matrix to diagonalise $M_R$, i.e.,
\begin{eqnarray} \label{eq:M_R}
    U_R^T M_R U_R = {\rm diag} \{ M_1, M_2, M_3 \} \,.
\end{eqnarray}

\item 
In the following analysis we take $M_R$  to be real, and  we assume that no CP violation is induced in the superpotential term ${\cal W}_{\nu^c}$. Then, $U_R$ is a real orthogonal matrix, parametrised by three angles as
\begin{eqnarray}
U_R = \left(\begin{matrix}
		c^{R}_{12} c^{R}_{13} & s^{R}_{12} c^{R}_{13} &
		s^{R}_{13} \cr \vspace{-0.4cm} \cr
		-s^{R}_{12} c^{R}_{23} - c^{R}_{12}
		s^{R}_{13} s^{R}_{23} & c^{R}_{12} c^{R}_{23} -
		s^{R}_{12} s^{R}_{13} s^{R}_{23}  & c^{R}_{13}
		s^{R}_{23} \cr \vspace{-0.4cm} \cr
		s^{R}_{12} s^{R}_{23} - c^{R}_{12} s^{R}_{13} c^{R}_{23}
		&- c^{R}_{12} s^{R}_{23} - s^{R}_{12} s^{R}_{13}
		c^{R}_{23} &  c^{R}_{13} c^{R}_{23} \cr
	\end{matrix} \right)
\end{eqnarray}
where $c_{ij}^R = \cos \theta^R_{ij}$ and $s_{ij}^R = \sin \theta^R_{ij}$. 
\end{itemize}
We discuss the flavour texture of $M_{\nu}$ which can be compatible with the current oscillation data. 

We first check in the simplified case with vanishing $\delta_M$ and $\kappa$. 
{Using the seesaw formula $M_{\nu}^0 = Y_\nu (M_R^{-1})^0 Y_\nu^T v_u^2$ where $(M_R^{-1})^0 = M_R^{-1}\big|_{\delta_M = \kappa = 0}$ is denoted and $Y_\nu = \hat{Y}_u$ is considered, we obtain
{\begin{eqnarray} \label{eq:M_nu_0}
M_{\nu}^0 \!= \!\!
\frac{v_u^2}{M}\left(
\begin{array}{ccc}
 y_u^2 (U_{11}^2 + U_{12}^2) & \!\!y_u y_c (U_{11} U_{21} + U_{12} U_{22})\!\! & \!\!y_u y_t (U_{11} U_{31} + U_{12} U_{32})\!\! \\
\!\! y_u y_c (U_{11} U_{21} + U_{12} U_{22})\!\! & y_c^2 (U_{21}^2 + U_{22}^2) & \!\!y_c y_t (U_{21} U_{31} + U_{22} U_{32})\!\! \\
\!\! y_u y_t (U_{11} U_{31} + U_{12} U_{32})\!\! & \!\!y_u y_t (U_{21} U_{31} + U_{22} U_{32})\!\! & y_t^2 (U_{31}^2 + U_{32}^2) \\
\end{array}
\right) \,,
\end{eqnarray}
}where $U_{ij}$ is the abbreviation of the $(i,j)$ entry of $U_R$, i.e., $U_{R,ij}$. 
Since $\det M_\nu^0 \propto \det (M_R^{-1})^0 = 0$, there will always be one eigenvalue of $M_\nu$ vanishing. For the two non-vanishing masses, naively, if all entries of $U_R$ are assumed to be domestically distributed, one can predict them proportional to $y_c^2$ and $y_t^2$, respectively. They are too hierarchical and conflict with neutrino oscillation data. To solve this problem, we will assume a special structure in $M_\nu^0$ as below.  
With the help of orthogonal condition $U_{R,i1} U_{R,j1}+ U_{R,i2} U_{R,j2} = \delta_{ij}-U_{R,i3} U_{R,j3}$, we re-write $M_{\nu}^0$ in the form
\begin{eqnarray} \label{eq:M_nu_0}
M_{\nu}^0 = 
\frac{v_u^2}{M}\left(
\begin{array}{ccc}
 y_u^2 & 0 & 0 \\
 0 & y_c^2 & 0 \\
 0 & 0 & y_t^2 \\
\end{array}
\right)
-\frac{v_u^2}{M}\left(
\begin{array}{ccc}
 y_u^2 U_{R,13}^2 & y_u y_c U_{R,13} U_{R,23} & y_u y_t U_{R,13} U_{R,33} \\
 y_u y_c U_{R,13} U_{R,23} & y_c^2 U_{R,23}^2 & y_c y_t U_{R,23} U_{R,33} \\
 y_u y_t U_{R,13} U_{R,33} & y_c y_t U_{R,23} U_{R,33} & y_t^2 U_{R,33}^2 \\
\end{array}
\right) \,.
\end{eqnarray}
}To reproduce the correct mass hierarchy for light neutrinos, we must assume $U_{R,31}, U_{R,32} \simeq {\cal O} (y_c/y_t)$, i.e.,
\begin{eqnarray}
\theta^R_{13},~ \theta^R_{23} \simeq {\cal O} (y_c/y_t) \simeq {\cal O} (10^{-3})  \,.
\end{eqnarray}
It is convenient to introduce two ${\cal O}(1)$ parameters
\begin{eqnarray} \label{eq:definition_a_b}
a = \Big(\frac{y_t}{y_c} \sin \theta_{13}^R\Big)^2\,,\quad 
b = \Big(\frac{y_t}{y_c} \sin \theta_{23}^R\Big)^2\,.
\end{eqnarray}
Then, the size of each entry of $M_\nu^0$ is estimated to be
\begin{eqnarray}
M_{\nu}^0 \simeq
\frac{m_c^2}{M}\left(
\begin{array}{ccc}
 {\cal O}(10^{-6}) &  {\cal O}(10^{-9}) & {\cal O}(10^{-3}) \\
  {\cal O}(10^{-9}) & 1 & \sqrt{b} \\
 {\cal O}(10^{-3})  & \sqrt{b} & a+b \\
\end{array}
\right) \,,
\end{eqnarray}
where $m_c = y_c v_u$. We refer to Eq.~\eqref{eq:M_nu_0_v2} in the appendix for the detailed expression of $M_\nu^0$. Then, assuming that the LH neutrino mass spectrum takes  normal hierarchy (NH), the three neutrino eigenmasses are given by 
\begin{eqnarray} \label{eq:m123_0}
&&m_1 = 0,\quad m_{2,3} \simeq \frac{m_c^2}{2M} \left[ 1+a+b \mp \sqrt{(1+a+b)^2-4a} \right]~.
\end{eqnarray}
 In the inverted hierarchy (IH), the replacement $(m_1, m_2, m_3) \to (m_3, m_1, m_2)$ is understood and will not be repeated in the following. 
We further find  that  $a$ and $b$ are related via the equation 
\begin{eqnarray} \label{eq:correlation_a_b}
b \simeq \sqrt{a} \left( \sqrt{\frac{m_2}{m_3}} + \sqrt{\frac{m_3}{m_2}} \right) - a -1
\end{eqnarray}
while the following restrictions hold on $a$ and $b$
\begin{eqnarray}
\frac{m_2}{m_3} \lesssim a \lesssim \frac{m_3}{m_2}\,, \quad
0 \leqslant b \lesssim \frac14 \left(\sqrt{\frac{m_3}{m_2}} + \sqrt{\frac{m_2}{m_3}} -2 \right) \,,
\end{eqnarray}
where the maximal value of $b$ is taken at $a = \frac14 (\frac{m_3}{m_2} + \frac{m_2}{m_3} + 2)^2$. As $m_1 =0$, we can take $m_2= \sqrt{\Delta m^2_{21}}$ and $m_3= \sqrt{\Delta m^2_{31}}$ explicitly. The unitary matrix which diagonalises $M_\nu$ is approximately expressed as
\begin{eqnarray}
U_\nu \simeq \left(
\begin{array}{ccc}
 1 &  0 & 0 \\
 0 & \cos\theta & \sin\theta \\
 0  & -\sin\theta & \cos\theta \\
\end{array}
\right) \label{eq:U_nu}
\end{eqnarray}
with 
\begin{eqnarray}
\sin2\theta = 2 \left(\sqrt{\frac{m_3}{m_2}} - \sqrt{\frac{m_2}{m_3}}\right)^{-1} \sqrt{\frac{b}{a}}
\end{eqnarray}

Next, we include the correction induced by the parameters $\delta_M$ and $\kappa$. $M_\nu  = M_\nu^0 + \delta M_\nu$ and $\delta M_\nu$ is estimated to be
\begin{eqnarray} \label{eq:M_nu_approx}
\delta M_{\nu} \simeq
\frac{m_c^2}{M}\left(
\begin{array}{ccc}
 {\cal O}(10^{-6}) \delta_M &  {\cal O}(10^{-3}) \delta_M & {\cal O}(10^{-3}) \delta_M + {\cal O}(10^{-3}) \kappa \\
  {\cal O}(10^{-3}) \delta_M & {\cal O}(1) \delta_M & {\cal O}(1) \delta_M + {\cal O}(1) \kappa \\
 {\cal O}(10^{-3}) \delta_M + {\cal O}(10^{-3}) \kappa & {\cal O}(1) \delta_M + {\cal O}(1) \kappa & \kappa \, y_t^2 / y_c^2 \\
\end{array}
\right) \,.
\end{eqnarray}
Eq.~\eqref{eq:delta_M_nu} in the appendix gives the detailed expression of $\delta M_\nu$. Most entries of $\delta M_\nu$ induce only small corrections to the neutrino masses and flavour mixing due to the suppression of $\delta_M$ and $\kappa$. The only exception is the $(3,3)$ entry, which is enhanced by $y_t^2 / y_c^2$. Including this term  leads to the modification of masses $m_2$ and $m_3$ by simply replacing $a$ with $a' = a+ \kappa y_t^2 / y_c^2$. A non-zero $\kappa$ also leads to a non-zero lightest left-handed (LH) neutrino mass. All light neutrino mass eigenvalues are approximately given by 
\begin{eqnarray} \label{eq:m123}
m_1 &\simeq& \frac{m_c^2}{M} \frac{y_t^2 \kappa}{y_c^2 a'} \frac{y_u^2}{y_c^2} \,,\nonumber\\ 
m_{2,3} &\simeq& \frac{m_c^2}{2M} \left[ 1+a'+b \mp \sqrt{(1+a'+b)^2-4a'} \right]
\end{eqnarray}
Again,
$a'$ and $b$ satisfy the relation 
\begin{eqnarray} \label{eq:correlation_ap_b}
b \simeq \sqrt{a'} \left( \sqrt{\frac{m_2}{m_3}} + \sqrt{\frac{m_3}{m_2}} \right) - a' -1
\end{eqnarray}
and the restriction
\begin{eqnarray} \label{eq:ab}
\frac{m_2}{m_3} \lesssim a' \lesssim \frac{m_3}{m_2}\,, \quad
0 \leqslant b \lesssim \frac14 \left(\sqrt{\frac{m_3}{m_2}} + \sqrt{\frac{m_2}{m_3}} -2 \right) \,.
\end{eqnarray}
The parameter $\kappa$, referring to the hierarchy between the heaviest RH neutrino mass and the other two, has to be very small, $\kappa = (a' -a) y_c^2 / y_t^2 \lesssim {\cal O}(10^{-6})$. 

The existence of the non-zero $\kappa$ has two main implications  on the LH neutrino masses and mixing.
The first one is to give a tiny mass to the lightest LH neutrino. From Eq.~\eqref{eq:m123}, we see that $m_1 \lesssim \frac{y_u^2}{y_c^2} m_{2,3} \sim 10^{-6} m_{2,3}$ since $\frac{y_t^2 \kappa}{y_c^2 a'} \lesssim {\cal O}(1)$ is required. We have numerically checked that $m_1 \lesssim 10^{-7}~{\rm eV}$. Thus, we can still take $m_2 \approx \sqrt{\Delta m_{21}^2}$ and $m_3 \approx \sqrt{\Delta m_{31}^2}$ approximately. 
The second implication is to modify the relation between $m_2$ and $m_3$ from Eq.~\eqref{eq:correlation_a_b} to  Eq.~\eqref{eq:correlation_ap_b}. The factor $\kappa y_t^2/y_c^2$ allows $\kappa$ to have an important contribution to the masses $m_2$ and $m_3$ even if $\kappa$ is tiny.
The other parameter $\epsilon$, referring to the mass splitting between $M_1$ and $M_2$, has a negligible effect on $M_\nu$, however,  is crucial for enhancing the CP asymmetry in leptogenesis. The unitary matrix $U_\nu$ has approximately  the same form as that in Eq.~\eqref{eq:U_nu} but with $\theta$ replaced by
\begin{eqnarray}
\sin2\theta = 2 \left(\sqrt{\frac{m_3}{m_2}} - \sqrt{\frac{m_2}{m_3}}\right)^{-1} \sqrt{\frac{b}{a'}}
\end{eqnarray}

\begin{figure}[h!]
\centering
\includegraphics[width=.45\textwidth]{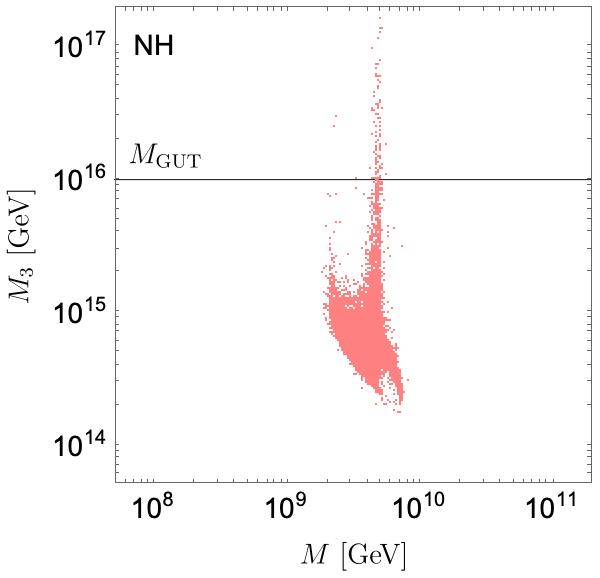}
\includegraphics[width=.45\textwidth]{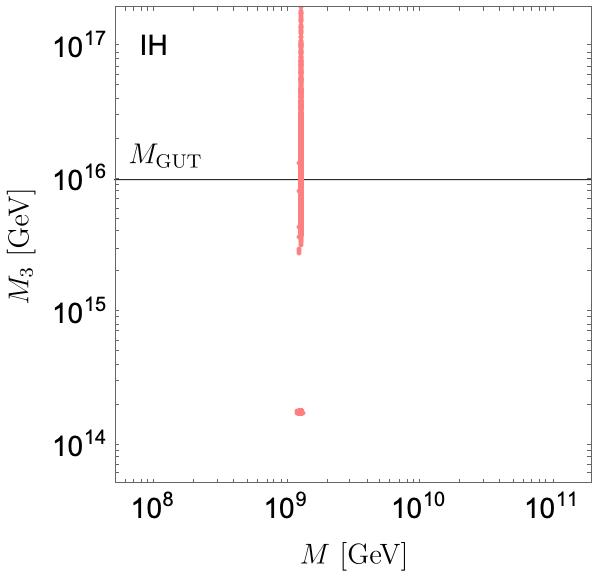}
\caption{Masses between $M \equiv (M_1+M_2)/2$ and $M_3$ in NH (left panel) and IH (right panel) cases, with $\delta_M \equiv (M_2-M_1)/(2M)$ logarithmically scanned in the range $(10^{-6}, 1)$.} 
\label{fig:RHN_mass}
\end{figure}

By means of this analytical discussion, we are able to perform a very efficient numerical scan by varying the input parameters in the derived intervals. 
In the numerical scan, we follow the subsequent  procedures: 
\begin{itemize}
    \item[1)]  $a'$ is treated as a free parameter in the interval shown in Eq.~\eqref{eq:ab}, $\kappa$ varies logarithmically in the interval $[10^{-6}, a'] y_c^2 / y_t^2$, and $a$ and $b$ are respectively determined by $a = a'- \kappa y_t^2 /y_c^2$ and Eq.~\eqref{eq:correlation_ap_b}. 
Once $a$ and $b$ are obtained, $\theta_{13}^R$ and $\theta_{23}^R$ are determined via Eq.~\eqref{eq:definition_a_b}. Here, all quark Yukawa couplings and mixing parameters are fixed at their best-fit values after RG running to the GUT scale\cite{Charalampous:2021gmf,King:2023wkm}. 
\item[2)] The third angle $\theta_{12}^R$ is assumed to vary randomly in the interval $(0, 2\pi)$. Once these parameters are introduced as inputs, $M_R$ is obtained via Eq.~\eqref{eq:M_R} up to an overall mass scale.
\item[3)] Through the seesaw formula, $M_\nu$ is derived also up to an overall mass scale and $U_\nu$ is calculated by the equation~\eqref{eq:master}. 
\item[4)]We do a simple $\chi^2$ analysis where $\chi^2 < 10$ values are considered, and  experimental data of three lepton mixing angles and two mass-squared differences are taken into account. Free input parameters include: $(a', \kappa, \theta_{12}^R)$ which are discussed in the above items, one overall mass scale for light neutrino, all oscillation parameters $(\theta_{12}, \theta_{13}, \theta_{23}, \delta)$ in the PMNS matrix which we assume to vary in their $1\sigma$ region, two Majorana phases $(\alpha_{21},\alpha_{31})$, and three phases $(\beta_1, \beta_2, \beta_3)$ in $P_l$.
\end{itemize}

In Fig.~\ref{fig:RHN_mass}, we show the correlation between the heaviest RH neutrino mass $M_3$ and the average  mass $M=(M_1+M_2)/2$ of two lighter RH neutrinos. 
The lower bound of $M_3$, which is close to the canonical seesaw scale $\sim 10^{14}$~GeV, is given by $\kappa y_t^2/y_c^2 \approx a'$ and $a\to 0$.  The upper bound of $M_3$ refers to $\kappa y_t^2/y_c^2 \to 0$ and $a\approx a'$. A cutoff for $M_3$ should be included as it is higher that the GUT scale $ M_{\rm GUT}\sim 10^{16}$~GeV. The IH case gives   robust prediction for RH neutrino masses, $M \approx 1.3 \times 10^9$~GeV and $M_3 \approx 1.8 \times 10^{14}$ or $\gtrsim 2.5 \times 10^{15}$~GeV. We explain these results below. Recall Eqs.~\eqref{eq:m123} and \eqref{eq:ab} with $m_1, m_2, m_3$ replaced by $m_3, m_1, m_2$ in the IH case. As $m_3$ almost vanishes, $m_1 \approx \sqrt{-\Delta m^2_{32}-\Delta m^2_{21}}$ and $m_2 \approx \sqrt{-\Delta m^2_{32}}$, the parameter $a'$, which is restricted in the range $[m_1/m_2, m_2/m_1]$, has to take a value very close to one, and $b \approx 0$. Then, we obtain $m_{1,2}\simeq m_c^2/M$, leading to the very restricted prediction for $M$. As for $M_3$, the two separated regions refer to $\kappa y_t^2/y_c^2 \approx a' \approx 1$ and $\kappa y_t^2/y_c^2 \ll 1$, respectively.

Note that our numerical scan is performed explicitly without any approximation. The analytical formulae  between $a$, $b$, $\kappa$, which are considered  above, are treated as a guideline to restrict the parameter space. 
In particular, they can be used to select points that give successful leptogenesis,
as  discussed in the next section. We emphasise that once $M_R$ is obtained, no approximation is used to derive the observables. 

\section{Leptogenesis} \label{sec:leptogenesis}

After having studied the flavour pattern of the neutrino sector, in this section we try to determine whether we can have successful leptogenesis for some portion of the parameter space. We denote the mass eigenstates for RH neutrinos as $N_1$, $N_2$ and $N_3$. 
We concentrate on the portion of the parameter space for which we have $M_1 \sim M_2$ and where $M_3$ is very  large and thus the contribution of the heaviest RH neutrino gets completely washed out. We will include only $N_1$ and $N_2$ in the evolution of leptogenesis. 

\begin{table}[h!]
\centering
\begin{tabular}{|c|c|c|}
\hline
& $M_2 - M_1$ (GeV) & $\Gamma$ (GeV) \\
\hline
BP1& $1.5 \times 10^{4}$ & $52$  \\
BP2 & $2.5 \times 10^{7}$ & $1.18$ \\
BP3 & $4.1 \times 10^{8}$ & $32$  \\
\hline
\end{tabular}
\begin{tabular}{|c|c|c|}
\hline
& $M_2 - M_1$ (GeV) & $\Gamma$ (GeV) \\
\hline
BP4& $1.1 \times 10^{4}$ & $118 $ \\
BP5 & $1.7 \times 10^{7} $& $440 $\\
BP6 & $6.9 \times 10^{8} $& $99$ \\
\hline
\end{tabular}
\caption{For resonant leptogenesis  $M_2-M_1 \simeq \Gamma$  is required. Above, in the left and right table  are the values for three benchmark points with different values of $\Delta_M$ respectively for the NH and IH scenarios. One can see that all of the points are outside the resonant condition. }
\label{tab:resNH}
\end{table}

\begin{figure}[t!]
\centering
\includegraphics[width=.95\textwidth]{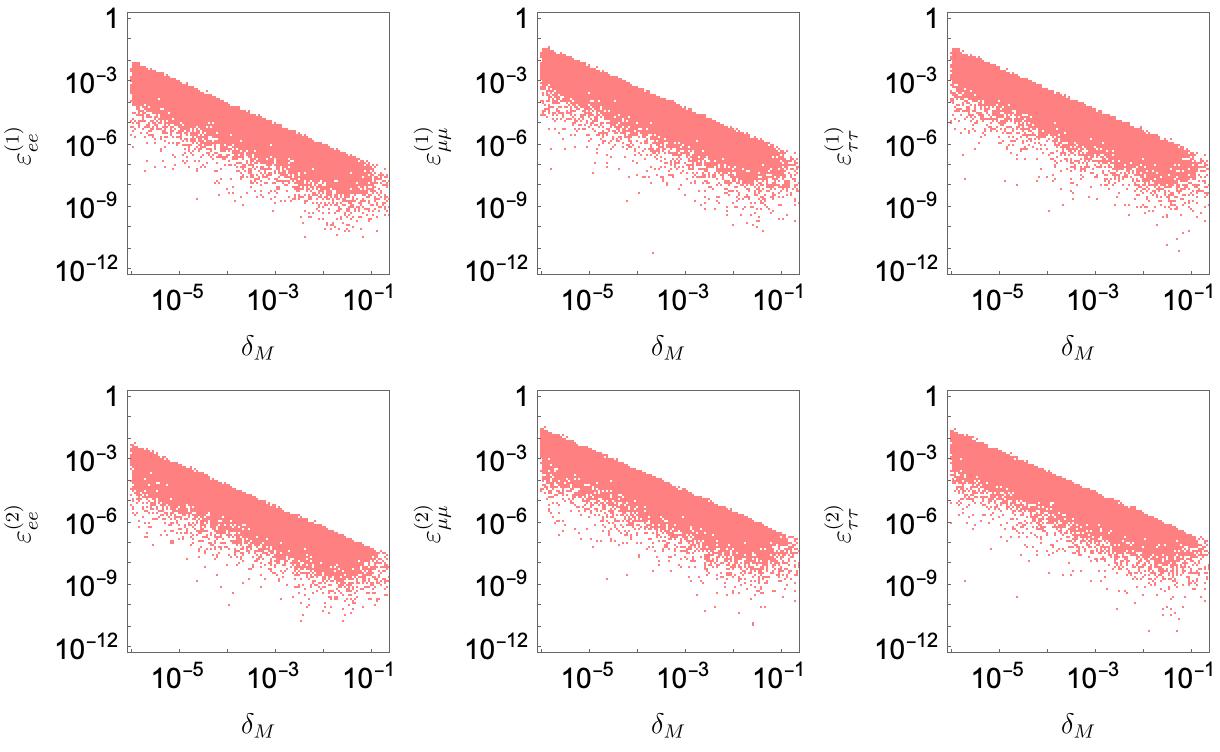}
\caption{The CP asymmetry for decay between $N_i \to h_u L_\alpha$ and its CP conjugate process with NH is assumed for illustration.}
\label{fig:epsilon}
\end{figure}

We use the density matrix formalism to calculate the baryon-antibaryon asymmetry. The density matrix equation for the asymmetry of the flavour $B-L$, (i.e., $B/3 - L_\alpha$) is~\cite{Blanchet:2011xq,Moffat:2018wke}
\begin{align}
\begin{aligned} & \frac{d N_{\alpha \beta}^{B-L}}{d z}=\sum_{i=1}^2 \varepsilon_{\alpha \beta}^{(i)} D_i\left(N_{N_i}-N_{N_i}^{\mathrm{eq}}\right)-\frac{1}{2} W_i\left\{\mathcal{P}^{(i) 0}, N^{B-L}\right\}_{\alpha \beta} \\ & -\frac{\operatorname{Im}\left(\Lambda_\tau\right)}{H z}\left[\left(\begin{array}{lll}1 & 0 & 0 \\ 0 & 0  & 0 \\ 0 & 0 & 0\end{array}\right),\left[\left(\begin{array}{lll}1 & 0 & 0 \\ 0 & 0 & 0 \\ 0 & 0 & 0\end{array}\right), N^{B-L}\right]\right]_{\alpha \beta} \\ & -\frac{\operatorname{Im}\left(\Lambda_\mu\right)}{H z}\left[\left(\begin{array}{lll}0 & 0 & 0 \\ 0 & 1 & 0 \\ 0 & 0 & 0\end{array}\right),\left[\left(\begin{array}{lll}0 & 0 & 0 \\ 0 & 1 & 0 \\ 0 & 0 & 0\end{array}\right), N^{B-L}\right]\right]_{\alpha \beta} \\ &
 \end{aligned}
\end{align}
Here, $\varepsilon^{(i)}_{\alpha \beta}$ is the CP source term including both vertex and self-energy contributions \cite{Covi:1996wh}

\begin{eqnarray} \label{eq:epsilon}
\varepsilon^{(i)}_{\alpha \beta} = \frac{1}{8 \pi (\tilde{Y}^\dag \tilde{Y})_{ii}} \sum_{j \neq i} 
{\rm Im} \left[ 
\tilde{Y}_{\alpha j} \tilde{Y}^*_{\beta i} (\tilde{Y}^\dag \tilde{Y})_{ij}
\right] g\Big(\frac{M_j^2}{M_i^2}\Big) +
{\rm Im}\left[ \tilde{Y}_{\alpha j} \tilde{Y}^*_{\beta i} (\tilde{Y}^\dag \tilde{Y})_{ji}
\right] f\Big(\frac{M_j^2}{M_i^2}\Big) \,,
\end{eqnarray}
where $\tilde{Y}$ is the Dirac neutrino Yukawa coupling matrix in the mass basis of charged leptons and RH neutrinos
\begin{eqnarray} \label{eq:Ytilde}
\tilde{Y} = U_l^\dag Y_\nu U_R =  U_{\rm PMNS} U_\nu^\dag \hat{Y}_u U_R \,,
\end{eqnarray} 
and the functions $g$ and $f$ are given by
\begin{eqnarray}
g (x) &=& \sqrt{x} \left[(x+1) \log \left(\frac{x+1}{x} \right) - \frac{2-x}{1-x} \right] \,, \nonumber\\
f (x) &=& \frac{1}{x-1}\,.
\end{eqnarray}
Since we are focusing on the nearly-degenerate case, then  $M_2-M_1 \ll M_1+M_2$. It is worth mentioning  that the above formula does not hold in the resonant case $M_i -M_j \lesssim \Gamma_i$, i.e., $\delta_M \lesssim (\tilde{Y}^\dag \tilde{Y})_{ii}/(4\pi)$ \cite{Pilaftsis:2003gt, DeSimone:2007edo}. 
Fortunately, as we are going to discuss below, in the parameter space we considered we have always $M_2-M_1 \gg \Gamma_{1,2} $, and thus we are safe to use just the density matrix formalism with the CP source in Eq.~\eqref{eq:epsilon}. 
In Fig.~\ref{fig:epsilon} we show plots  of the CP asymmetry $\epsilon^{(i)}_{\alpha\alpha}$ for the decay $N_i \to h_u L_\alpha$ and its CP conjugate process. 
{Considering we have assumed a real matrix $U_R$, it is straightforward to show that $\tilde{Y}^\dag \tilde{Y} = U_R^T \hat{Y}_u^2 U_R$ is real and symmetric.  We can further derive the analytical formula of $\epsilon_{\alpha\alpha}^{(i)}$ approximately in the nearly-degenerate but non-resonant regime for RHN neutrino masses, i.e., $\Gamma \ll M_2- M_1 \ll M_2+M_1$. And they are given by
\begin{eqnarray}
    \epsilon_{\alpha\alpha}^{(1)} \simeq
    \frac{(\tilde{Y}^\dag \tilde{Y})_{12}}{8 \pi (\tilde{Y}^\dag \tilde{Y})_{11}} \,
{\rm Im}  \left( 
\tilde{Y}_{\alpha 2} \tilde{Y}^*_{\alpha 1} 
\right) \, \frac{M_2+M_1}{M_2-M_1} \,, \quad
\epsilon_{\alpha\alpha}^{(2)} \simeq 
\epsilon_{\alpha\alpha}^{(1)} \frac{(\tilde{Y}^\dag \tilde{Y})_{11}}{(\tilde{Y}^\dag \tilde{Y})_{22}} \,.
\end{eqnarray}
Recall Eq.~\eqref{eq:Ytilde} that the only CP-violating phases in $\tilde{Y}$ is from the PMNS matrix. Thus, we obtain a direct connection between CP violation in the heavy RHN neutrino decay and that in light neutrino experiments. However, this result is based on the assumption of a real $U_R$, which is crucial in deriving the flavour pattern as well as the correlation between light and heavy neutrino masses in the last section. Without this assumption, i.e., adding some phases to $U_R$, we will lose the prediction.  
}

\begin{figure}[t!]
\centering
\includegraphics[width=.95\textwidth]{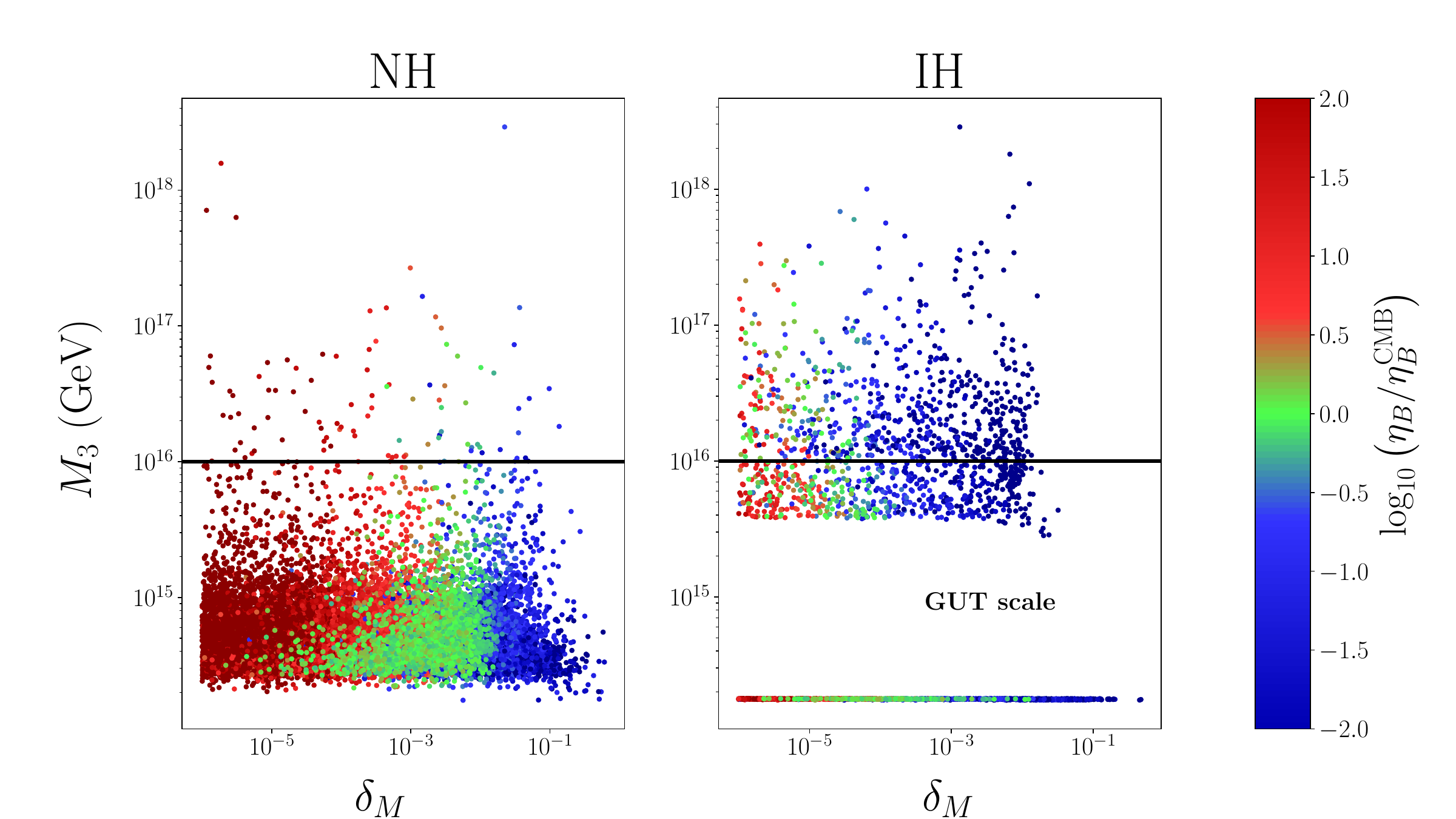}
\caption{The dependence of the baryon-antibaryon asymmetry $\eta_b$ on $\delta_M$ and $M_3$ in both NH (left panel) and IH (right panel) cases. On green are the points for which $\eta_b/\eta_b^{\mathrm{BBN}}$ is of order $\mathcal{O}(1)$. 
The black solid line refers to the GUT scale around $10^{16}$~GeV. The RH neutrino mass $M_3$ above this line has the non-perturbative problem and should not be considered.}
\label{fig:mass_scales}
\end{figure}
\begin{figure}[h!]
\centering
\includegraphics[width=.95\textwidth]{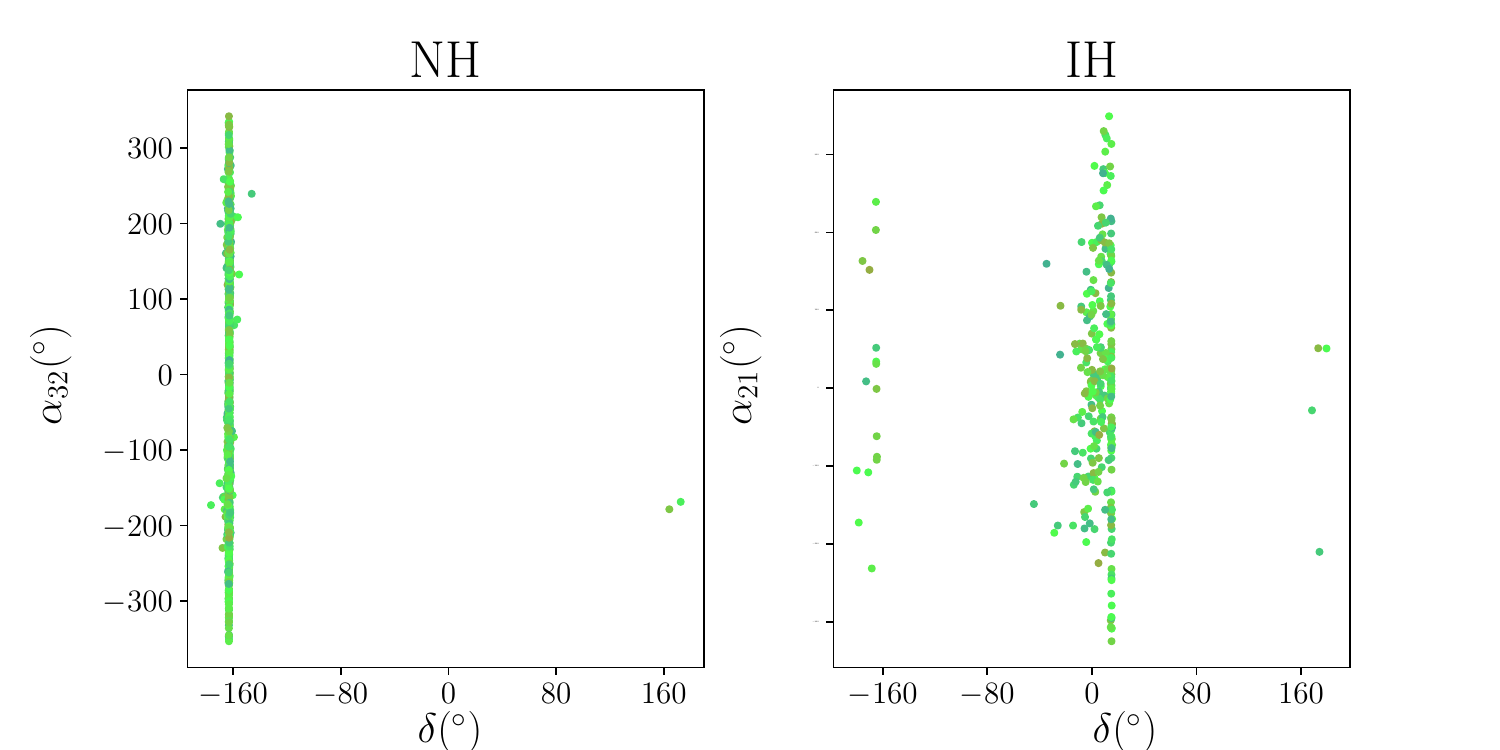}
\caption{The dependence of $\eta_b$ on $\delta$ and $\alpha_{32} = \alpha_{31} - \alpha_{21}$ in the NH case (left panel), and $\delta$ and $\alpha_{21}$ in the IH case (right panel).  Only  points within the 3$\sigma$ range (in green) are shown. }
\label{fig:phases}
\end{figure}

\begin{figure}[h!]
\centering
\includegraphics[width=.95\textwidth]{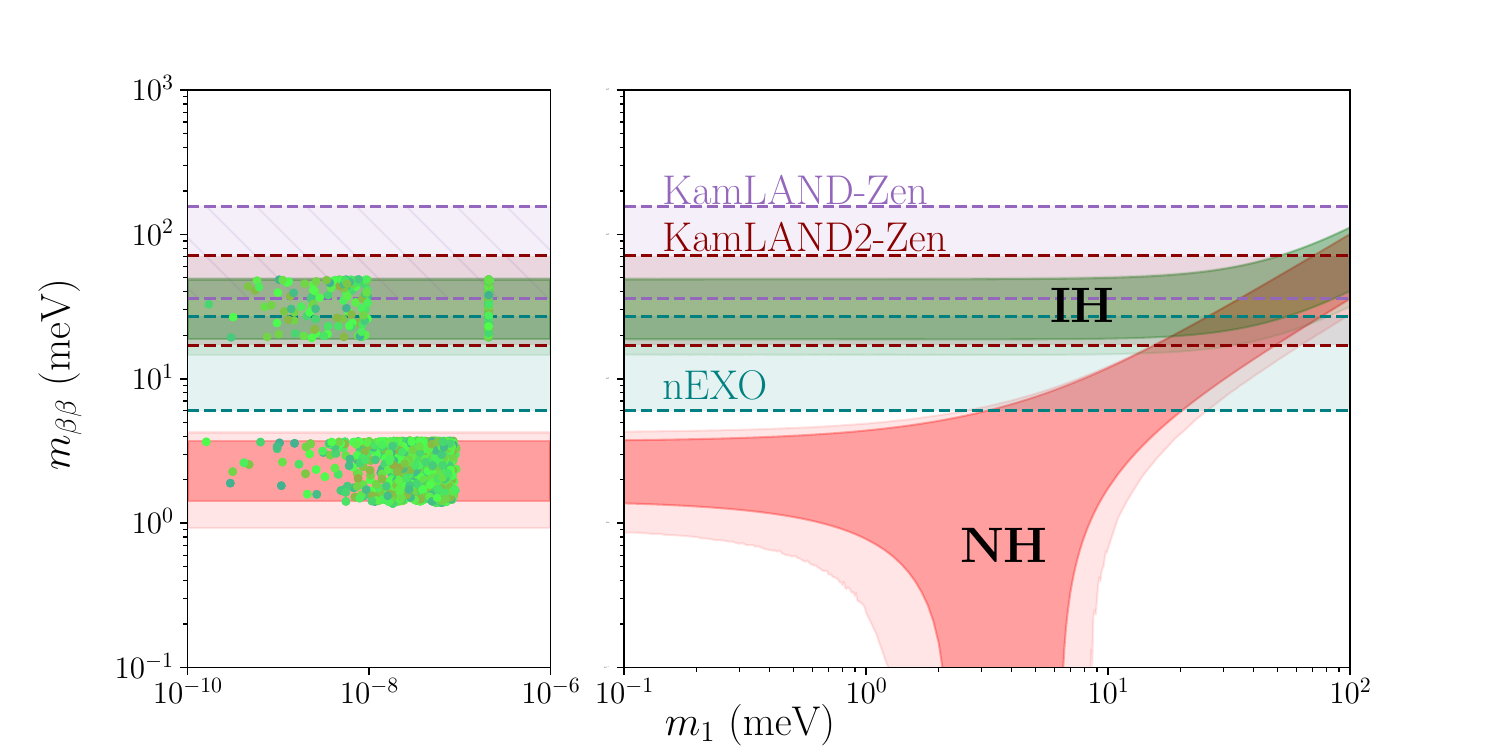}
\caption{$m_{ee}$ against the lightest active neutrino mass $m_1$ for NH  and IH. $m_1$ is predicted to be very light in both  scenarios and the results are showed in the left panel. 
Only points within the 3$\sigma$ range (in green) are shown. Current best experimental limit from KamLAND-Zen \cite{KamLAND-Zen:2022tow} and future sensitivities in KamLAND2-Zen and nEXO  \cite{Agostini:2022zub} are shown.}
\label{fig:Majorana}
\end{figure}
We apply Universal LeptogeneSiS Equation Solver ({\tt ULYSSES}) \cite{Granelli:2020pim, Granelli:2023vcm} in our numerical calculation of leptogenesis. We have used  2-flavour density matrix equation (2DME)
and resonant leptogenesis (2RES)
formulations to calculate the baryon asymmetry $\eta_B$
for a cross check. 
In general, results in both methods are consistent with each other if the mass splitting is not too small. Indeed, the Yukawa couplings with $N_1$ and $N_2$ are in general of order $10^{-3}$, and the resonant region,  i.e., $M_2 - M_1 \lesssim \Gamma_{1,2}$, appears only for $\delta_M \lesssim (\tilde{Y}^\dag \tilde{Y})_{ii}/(4\pi) \sim 10^{-7}$. We have checked that this region gives a huge  value for $\eta_B$, and 
thus the 2DME formulation is enough for us to do the scan. 
In Fig.~\ref{fig:mass_scales}, we show the prediction of $\eta_B$ as a function of $\delta_M$ and $M_3$ with 2DME applied. 
        
        
We found out that there is a region of the parameter space for which we achieve successful leptogenesis. There is a strong correlation between the quantity $\delta_M$ and the predicted $\eta_b$ as one can see from Fig.~\ref{fig:mass_scales}. 
Moreover, in Fig~\ref{fig:phases} we notice that in this model there is a sharp prediction for the Dirac phase; in the NH case it is far from the maximum CP violating case but it still contributes to generate the lepton asymmetry, while in the IH scenario the major contribution to the asymmetry is always given by the Majorana phase. 
Finally, in Fig.~\ref{fig:Majorana} we show the predictions for $m_{e e}$ in comparison with the mass of the lightest active neutrino $m_1$. The IH scenario can be tested with future $0 \nu \beta \beta$ experiments 
\cite{Agostini:2022zub}, 
while the NH cannot. For both scenarios due to what we discussed above, $m_1$ is very small, several orders of magnitude smaller than the reach of next generations laboratory and cosmological experiments.

\section{Conclusion} \label{sec:conclusion}

Flipped SU(5) provides an attractive alternative  grand unified model to the well-known SU(5) and SO(10) GUTs.
A distinct feature of this model
is the prediction of very long proton lifetimes. Therefore, in the absence of proton decay in next-generation neutrino experiments, the model cannot be ruled out.
%
In the flavour space, this model predicts a correlation between Dirac neutrino and up-quark Yukawa couplings, thus it is natural to expect right-handed (RH) neutrinos to have a very hierarchical mass spectrum to fit the low-energy neutrino data via the seesaw mechanism. As a consequence, a successful leptogenesis may be hard to achieve, since 1) the lightest RH neutrino is too light to generate enough baryon-antibaryon asymmetry and 2) any baryon-antibaryon asymmetry generated by the heavier RH neutrinos might be washed out by the lightest one, unless special flavour textures are included to suppress the washout effect. 

In this paper, we  provided another option to apply  thermal leptogenesis to explain baryon-antibaryon asymmetry in the observed Universe. 
The key point for a successful leptogenesis in flipped SU(5) is the assumption that the two lighter RH neutrinos are approximately equal. We found, through analytical approximations and straightforward numerical calculations, that these two RH neutrinos have masses slightly above $10^9$~GeV and the heaviest one is between the classical seesaw scale $\sim 10^{14}$~GeV and the GUT scale $\gtrsim 10^{16}$~GeV.
As a consequence, the lightest left-handed neutrino, regardless of the normal or inverted hierarchy, has a very tiny mass $m_{\rm lightest}\lesssim 10^{-7}$~eV, though not exactly zero. The two lighter RH neutrino masses are heavy enough for thermal leptogenesis to apply. The small mass-splitting between them provides an enhancement of the CP asymmetry of RH neutrino decay. We found that the best region for the mass splitting should be around two to four orders of magnitude smaller than the mass scale. However this is not the resonant regime, which would require the mass splitting to be the same order as the decay width, and would overproduce the lepton asymmetry. For a normal neutrino mass hierarchy the model makes a sharp prediction for the CP violating Dirac phase with the bulk of the points in the range $\delta=160^\circ \sim 165^\circ$. 


\section*{Acknowledgements}

This work is supported by the STFC Consolidated Grant ST/L000296/1 and the European Union’s Horizon 2020 Research and Innovation programme under Marie Sklodowska-Curie grant agreement HIDDeN European ITN project (H2020-MSCA-ITN-2019//860881-HIDDeN) (S.F.K.), and National Natural Science Foundation of China (NSFC) under Grants Nos. 12205064, 12347103 and Zhejiang Provincial Natural Science Foundation of China under Grant No. LDQ24A050002 (Y.L.Z.).
L.M. is supported by the Spanish National Grant PID2022-137268NA-C55 and Generalitat Valenciana through the grant CIPROM/22/69.
S.F.K., G.L., and Y.L.Z. are grateful to the Mainz Institute for Theoretical Physics (MITP) of the Cluster of Excellence PRISMA+ (Project ID 390831469), for its hospitality and support. 

\appendix
\section{Derivation of leptogenesis-favoured flavour textures}
In this appendix, we show how to find the parameter space of fermion flavour structures in flipped SU(5) with our analytical approximation. This analytical approach is very helpful for us to restrict the parameter space in the multi-dimensional scan, and thus makes the numerical scan more efficient. However, since the analytical approach applies  only to  a specified region, we do not use it directly in the numerical calculations. We just get random points around the region suggested by the analytical approach and  do the explicit numerical calculation without any approximation. 

 The left-handed neutrino mass matrix, derived via  the seesaw formula, is given by
\begin{eqnarray}
    M_\nu = M_\nu^0 + \delta M_\nu
\end{eqnarray} with $M_\nu^0$ in Eq.~\eqref{eq:M_nu_0} and $\delta M_\nu$ given by 
\begin{eqnarray}
    \delta M_\nu = \frac{v_u^2}{M} Y_u 
U_R \left(
\begin{array}{ccc}
 \delta_M & 0 & 0 \\
 0 & -\delta_M & 0 \\
 0 & 0 & \kappa \\
\end{array}
\right) U_R^T
 Y_u \,.
\end{eqnarray}

It is obvious that ${\rm det}(M_\nu^0) = 0$ and thus one of the eigenvalues vanishes (we denote this eigenvalue as $\lambda_1^0$). The other two eigenvalues (denoted as $\lambda_2^0$ and $\lambda_3^0$) can be obtained by solving the equations
\begin{eqnarray}
    \lambda^0_2 + \lambda^0_3 &=& {\rm tr} (M_\nu^0) = 
   \frac{v_u^2}{M} \big[ (1-U_{R,13}^2) y_u^2 + (1-U_{R,23}^2) y_c^2 + (1-U_{R,33}^2) y_t^2 \big] \,, \nonumber\\
    \lambda^0_2 \lambda^0_3 &=& \frac12 \big[({\rm tr} M_\nu^0)^2 - {\rm tr} ((M_\nu^0)^2)\big] = 
    \frac{v_u^2}{M} \big[ U_{R,33}^2 y_u^2 y_c^2 + U_{R,23}^2 y_u^2 y_t^2 + U_{R,13}^2 y_c^2 y_t^2 \big] \,.
\end{eqnarray}
Here $\lambda_2^0$ and $\lambda_3^0$ give the two non-vanishing light neutrino masses, i.e., $m_2$ and $m_3$ in the NH. One cannot assume all $U_R$ entries $U_{R,13}$, $U_{R,23}$ and $U_{R,33}$ of order ${\cal O}(1)$, otherwise $\lambda^0_2 + \lambda^0_3 \sim y_t^2$ and $\lambda^0_2 \lambda^0_3 \sim y_c^2 y_t^2$, leading to very large hierarchical neutrino mass ratio $\sim y_c^2 /y_t^2$, which is inconsistent with the neutrino data. Instead, one has to assume $U_{R,13}^2 + U_{R,23}^2 = 1- U_{R,33}^2 \ll 1$. We consider the scenario 
\begin{eqnarray}
    U_{R,13} \sim U_{R,23} \sim \sqrt{1-U_{R,33}^2} \sim y_c / y_t \,.
\end{eqnarray}
In this case, we introduce the order-one parameters $a = U_{R,13}^2 y_t^2 / y_c^2$ and $b = U_{R,23}^2 y_t^2 /y_c^2$,\footnote{This is consistent with definitions in Eq.~\eqref{eq:definition_a_b} in the case of small $\theta_{13}^R$ and $\theta_{23}^R$.} which are helpful to simplify the analytical formulae. 
Then we obtain two eigenvalues as
\begin{eqnarray}
    \lambda_{2,3}^0 = \frac{m_c^2}{2 M}  \Big[1 + a + b \mp \frac{1}{2}\sqrt{(1 + a + b)^2 - 4 a} + {\cal O}(\frac{y_u}{y_c}) \Big] \,,
\end{eqnarray}
with $m_c = y_c v_u$. 
Ignoring terms suppressed by $y_u/y_c$, we obtain Eq.~\eqref{eq:m123_0} and the correlation in Eq.~\eqref{eq:correlation_a_b}. $M_\nu^0$ after introducing parameters $a$ and $b$ is written explicitly as
\begin{eqnarray} \label{eq:M_nu_0_v2}
    M_\nu^0 = \frac{m_c^2}{M} \begin{pmatrix}
      \frac{y_u^2}{y_c^2}-\frac{y_u^2}{y_t^2} a  & \frac{y_u y_c}{y_t^2}\sqrt{a b} & \frac{y_u}{y_c}\sqrt{a (1 - \frac{y_c^2}{y_t^2}(a+b))} \\ 
      \frac{y_u y_c}{y_t^2}\sqrt{a b} & 1 & -\sqrt{b} \\ 
      \frac{y_u}{y_c}\sqrt{a (1 - \frac{y_c^2}{y_t^2}(a+b))} & -\sqrt{b} & a+b
    \end{pmatrix}
\end{eqnarray}

We then include the contribution of $\delta M_\nu$. We write it in the form
\begin{eqnarray} \label{eq:delta_M_nu}
    \delta M_\nu = \frac{m_c^2}{M}\begin{pmatrix}
        \frac{y_u^2}{y_c^2} \delta_{11} & \frac{y_u}{y_c} \delta_{12} & \frac{y_u y_t}{y_c^2} \delta_{13} + \frac{y_u}{y_c} \sqrt{a} \kappa \\
        \frac{y_u}{y_c} \delta_{12} & \delta_{22} & \frac{y_t}{y_c} \delta_{23} + \sqrt{b}\kappa \\
        \frac{y_u y_t}{y_c^2} \delta_{13} + \frac{y_u}{y_c} \sqrt{a} \kappa & \frac{y_t}{y_c} \delta_{23} + \sqrt{b}\kappa & \frac{y_t^2}{y_c^2}\delta_{33} + \frac{y_t^2}{y_c^2} \kappa
    \end{pmatrix}
\end{eqnarray}
where $\delta_{ij} = (U_{R,i1} U_{R,j1} - U_{R,i2} U_{R,j2}) \delta_M$ refer to contributions from the mass splitting between $M_1$ and $M_2$.
This parametrisation is helpful for us to estimate the size of each entry of $\delta M_\nu$. We will not assume any  hierarchy among $U_{R,11}$, $U_{R,12}$ and $U_{R,21}$ and $U_{R,22}$. Namely these parameters can maximally reach order one, and thus $\delta_{11}, \delta_{12}, \delta_{22} \lesssim \delta_m$. $\delta_{13} = (U_{R,11}\sqrt{a}-U_{R,12} \sqrt{b})\frac{y_c}{y_t} \delta_m$ and $\delta_{23} = (U_{R,21}\sqrt{a}-U_{R,22} \sqrt{b})\frac{y_c}{y_t} \delta_m$, leading to $\frac{y_u y_t}{y_c^2}\delta_{13} \lesssim \frac{y_u}{y_c} \delta_m$ and $\frac{y_t}{y_c}\delta_{23} \lesssim \delta_m$. In the last entry, $\delta_{33} = \frac{y_c^2}{y_t^2}(a-b)\delta_m$, leading to $\frac{y_t^2}{y_c^2} \delta_{33} \lesssim \delta_m$. Thus, we estimated that the size of contribution of mass splitting between $M_1$ and $M_2$ to $\delta M_\nu$ can maximally reach the order $\delta_m$. In the preferred regime, as we discussed in the main text, the lightest two RH neutrinos are nearly degenerate, $\delta_m \ll 1$, contribution of $\delta_m$ does not have to be included in the analytical approximation. Eventually, we are left with a mainly contribution of $\kappa$, 
\begin{eqnarray}
    \delta M_\nu = \frac{m_c^2}{M} \left[\begin{pmatrix}
        0 & 0 & 0 \\ 0 & 0 & 0 \\ 0 & 0 & \frac{y_t^2}{y_c^2} \kappa
    \end{pmatrix} + {\cal O} (\delta_m) \right]
\end{eqnarray}
The $\kappa$ term is important if $\kappa \gtrsim y_c^2 /y_t^2$. Estimation of relative sizes for each entry of $\kappa$, or equivalently $M_\nu$, is summarised in Eq.~\eqref{eq:M_nu_approx}. Approximately, it has little difference, just only replacing $a$ by $a' = a + \kappa y_t^2 /y_c^2$. The eigenvalues of $\lambda_2$ and $\lambda_3$ can be approximately calculated by replaced $a$ by $a'$. The main difference between $M_\nu$ and $M^0_\nu$ is that the smallest eigenvalue $\lambda_1$ is no longer exactly zero, although still  highly suppressed by $y_u^2 /y_c^2 \simeq {\cal O}(10^{-6})$ following the analytical approximate solution in Eq.~\eqref{eq:m123}. Thus the lightest light neutrino mass $m_{\rm lightest}$ is six orders of magnitude lighter than the heaviest light neutrino mass, i.e., $m_{\rm lightest} \lesssim 10^{-7}$~eV.


\end{document}